# Ambipolar solution-processed hybrid perovskite phototransistors

Feng Li*, Chun Ma, Hong Wang, Weijin Hu, Weili Yu, Arif D. Sheikh & Tom Wu*

*Division of Physical Sciences and Engineering, King Abdullah University of Science and Technology, Thuwal 23955-6900, Kingdom of Saudi Arabia*

*E-mail: feng.li@kaust.edu.sa; tao.wu@kaust.edu.sa

## Abstract

**Organolead halide perovskites have attracted substantial attention due to their excellent physical properties, which enable them to serve as the active material in emerging hybrid solid-state solar cells. Here we investigate the phototransistors based on hybrid perovskite films, and provide direct evidence for their superior carrier transport property with ambipolar characteristics. The field-effect mobilities for triiodide perovskites at room temperature are measured as 0.18 (0.17) $cm^2 V^{-1} s^{-1}$ for holes (electrons), which increase to 1.24 (1.01) $cm^2 V^{-1} s^{-1}$ for mixed halide perovskites. The photoresponsivity of our hybrid perovskite devices reaches 320 $A W^{-1}$, which is among the largest values reported for phototransistors. Importantly, the phototransistors exhibit an ultrafast photoresponse speed of less than 10 μs. The solution-based process and excellent device performance strongly underscore hybrid perovskites as promising material candidates for photoelectronic applications.**

## Introduction

Methylammonium lead halide ($CH_3NH_3PbX_3$, X = halogen) perovskites have been intensively pursued as light-harvesting materials in high-performance hybrid solid-state photovoltaic devices[1-13]. This class of materials was first discovered by Weber nearly 36 years ago[14], and Mitzi and co-workers further revealed that halide perovskites combine the favourable carrier transport of inorganic semiconductors with the facile processing of organic materials[15]. In their pioneering work on field-effect transistors using hybrid perovskite $(C_6H_5C_2H_4NH_3)_2SnI_4$ as channels, a high On/Off ratio of $10^4$ and a hole mobility of 0.6 $cm^2 V^{-1} s^{-1}$ were reported[16]. The recent successes of halide perovskites in photovoltaic technologies can be primarily ascribed to their suitable, direct bandgap with large absorption coefficients, and low-cost solution-based processing, as well as their excellent transport properties[17-20]. Long electron-hole diffusion lengths and carrier lifetimes have also been observed in perovskite films, indicating low recombination rates of charge carriers[21,22]. The superb properties of these emerging materials also led to photoelectronic applications, such as electrically pumped lasers, light-emitting diode/transistors, and photodetectors[23-31]. Supplementary Table 1 summarizes the recent progresses on developing perovskite-based photodetectors. These hybrid perovskites can also be envisioned as good candidates for phototransistor, in which the gate bias, in addition to the incident light, is used as an additional parameter to modulate the channel transport[32,33]. Various types of semiconductors have been investigated as channel materials in phototransistors, such as Si, III-V semiconductors, ZnO, carbon nanotubes, quantum dots, organics and two-dimensional materials[34-49]. To date, there have been some pioneering works on utilizing $CH_3NH_3PbX_3$ as the active component in phototransistor-type devices[27,29,30]. However, the performance of these devices, particularly channel mobility and gate tuning, needs improvements.

In this work, we report on the fabrication and characterization of phototransistors based on solution-processed organolead triiodide ($CH_3NH_3PbI_3$) and mixed-halide ($CH_3NH_3PbI_{3-x}Cl_x$) perovskite films. The effects of gate voltage and light illumination on the transport of the perovskite channels were investigated. We found that these phototransistors exhibit clear ambipolar carrier transport characteristics, i.e., they work in both accumulation (p-type) and inversion (n-type) modes. Highly balanced photo-induced carrier mobilities of 0.18 $cm^2\ V^{-1}\ s^{-1}$ (holes) and 0.17 $cm^2\ V^{-1}\ s^{-1}$ (electrons) were observed for $CH_3NH_3PbI_3$, which increase to 1.24 $cm^2\ V^{-1}\ s^{-1}$ (holes) and 1.01 $cm^2\ V^{-1}\ s^{-1}$ (electrons) for the doped variant $CH_3NH_3PbI_{3-x}Cl_x$. As an important figure-of-merit for phototransistors, the photoresponsivity ($R$) of our perovskite phototransistor reaches 320 A $W^{-1}$. Furthermore, we observed that the phototransistors exhibit an ultrafast photoresponse time of less than 10 μs. The results indicate that solution-processed hybrid perovskites are very promising materials for constructing high-performance phototransistors, and they warrant exploration for use in other photoelectronic applications.

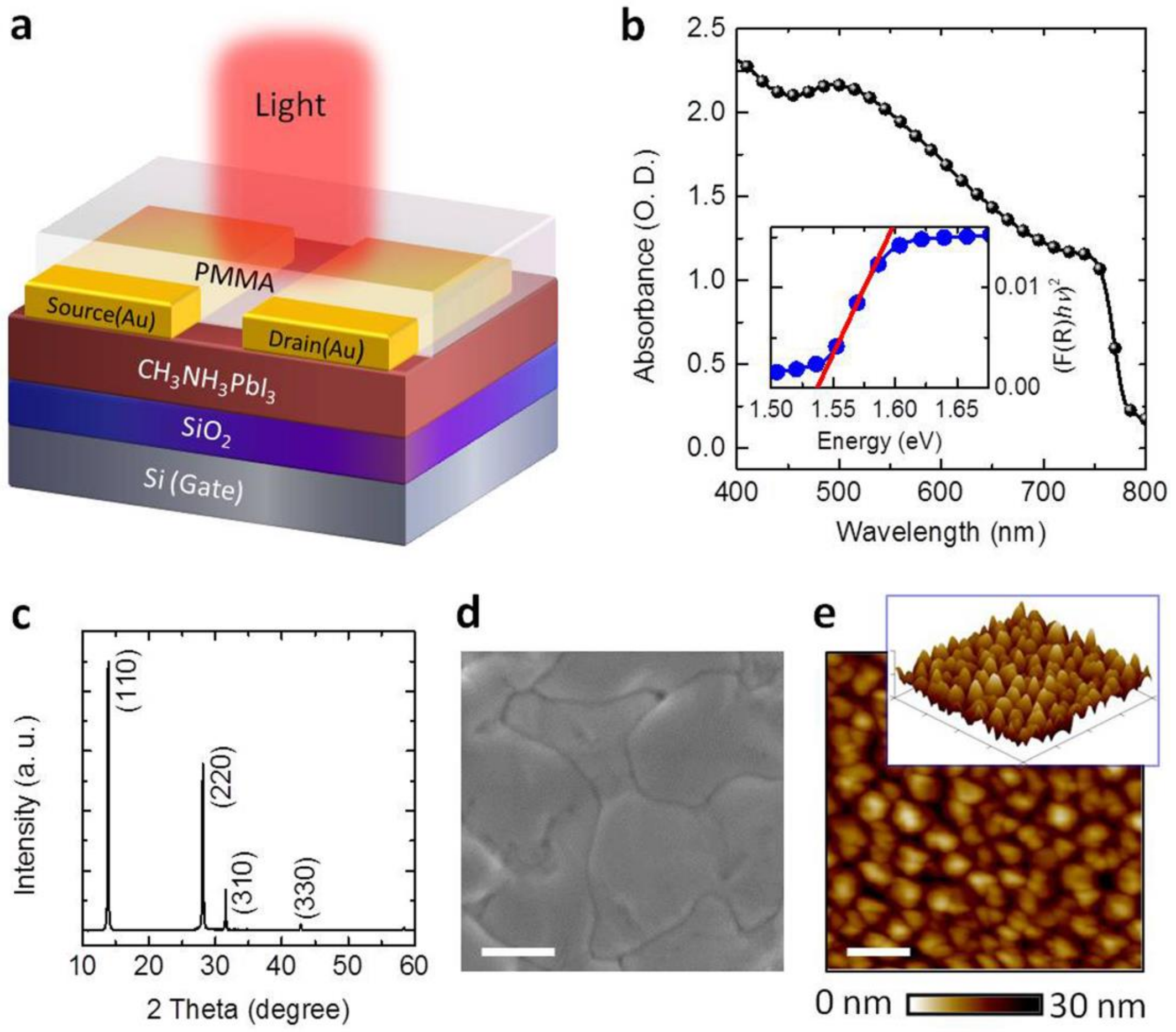

**Figure 1.** (**a**) Schematic of the phototransistor with a channel of hybrid perovskite $CH_3NH_3PbI_3$. (**b**) UV-visible absorption of the 100-nm $CH_3NH_3PbI_3$ film. The inset shows a direct band gap of approximately 1.53 eV. (**c**) X-ray diffraction spectrum of the $CH_3NH_3PbI_3$ film. (**d**) Top-view SEM image of the perovskite film. Scale bar: 0.5 μm. (**e**) Tapping-mode AFM height image (5.0 × 5.0 μm) of the perovskite film. Scale bar: 1.0 μm. Inset: the corresponding 3D topographic image.

A schematic device structure of a bottom-gate, top-contact hybrid perovskite $CH_3NH_3PbI_3$-based phototransistor is illustrated in Fig. 1a. A heavily n-doped Si wafer with a 300-nm $SiO_2$ surface layer (capacitance ($C_i$) of 15 nF cm$^{-2}$) was employed as the substrate. The perovskite films were grown using the two-step vapor-assisted solution process[11], and the experimental details are given in the Methods section. In our devices, we coated poly(methyl methacrylate) (PMMA) as the protective layer on top of the perovskite channel to prevent the diffusion of moisture and/or atmospheric oxygen[21]. In optimizing phototransistor devices, the semiconducting channel thickness is a key parameter. On the one hand, if the semiconducting channel is too thin, it will not absorb sufficient light[10]. In addition, pinholes in the thin perovskite films will cause inhomogeneous conduction in the channel. On the other hand, if the film is too thick, light from the top may not be able to penetrate the whole film, and the bottom gate will not effectively modulate the channel. In our work, the thickness of the $CH_3NH_3PbI_3$ film prepared in the phototransistors is optimized at 100 nm, as characterized using atomic force microscopy (AFM)[50]. The detailed AFM measurement results are shown in Supplementary Figure 1.

Figure 1b shows the light absorption data of the $CH_3NH_3PbI_3$ film. For this measurement, the thin films were prepared on a glass substrate using the same processing parameters as those used in fabricating the phototransistors. The strong and broad absorption in the UV and visible light range, particularly from 400 nm to 760 nm, reveals that the perovskite layer is a good light absorber. Furthermore, the absorption edge for the perovskite film is very sharp, suggesting a direct band gap nature[7,18]. Overall, the direct band gap of 1.53 eV and the favourable light absorption properties make perovskite $CH_3NH_3PbI_3$ films very promising materials for constructing high-performance phototransistors.

To investigate the crystallinity and microstructure of the perovskite thin films, X-ray diffraction (XRD) measurements were carried out. Figure 1c shows the XRD pattern of the as-prepared $CH_3NH_3PbI_3$ film on a $SiO_2$/Si substrate. Strong peaks at 14.08°, 28.41°, 31.85°, and 43.19° can be assigned to (110), (220), (310), and (330) diffractions of $CH_3NH_3PbI_3$, respectively, indicating that the halide perovskite films possess the expected orthorhombic crystal structure with high crystallinity[17]. Notably, the absence of a diffraction peak at 12.65° suggests that the level of the $PbI_2$ impurity phase is negligible[11]. The surfaces of the perovskite $CH_3NH_3PbI_3$ films were further evaluated via scanning electron microscopy (SEM) and atomic force microscopy (AFM) measurements. Figure 1d shows the surface morphology of the as-grown perovskite thin film grown on the $SiO_2$/Si substrate. The perovskite film appears smooth without pinholes and the uniform grains have sizes up to hundreds of nanometers. These structural characters are promising for achieving high performance in phototransistors. As shown in Fig. 1e, the film surface was also characterized using AFM, and the root-mean square roughness is approximately 10.5 nm in a typical scanning area of 5.0 μm × 5.0 μm. The three-dimensional AFM image in the inset of Fig. 1e further demonstrates the smooth surface of the perovskite film.

Figure 2a shows a representative set of the transfer characteristics, which are the drain current versus gate voltage ($I_{DS}$ - $V_{GS}$) data, of a bottom-gate top-contact $CH_3NH_3PbI_3$ phototransistor. The devices were measured both in the dark and under white light illumination (power density: 10 mW $cm^{-2}$) at the drain voltages ($V_{DS}$) of -30 V and 30 V. An ambipolar high performance can be clearly observed in the transfer characteristics under light illumination, and the V-shape of the transfer curves is similar to those of previous reports on ambipolar transistors[51-56]. The current of the illuminated channel can reach 0.1 mA. In contrast, for the transfer characteristics measured in the dark condition, $I_{DS}$ remains below 0.5 nA.

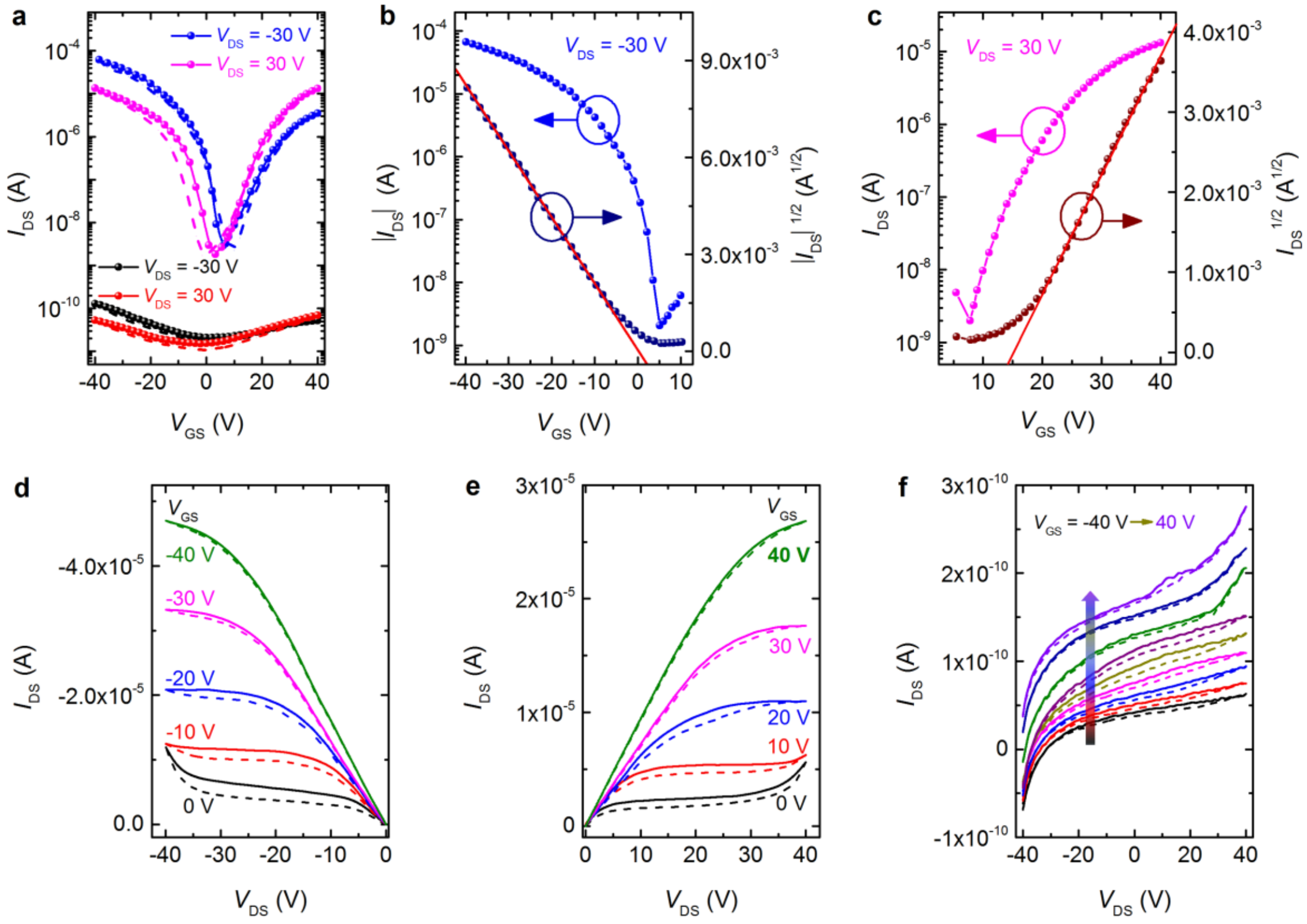

**Figure 2.** (**a**) Transfer characteristics of a perovskite-based phototransistor in the dark (black and red dots) and under light illumination (blue and magenta dots). (**b**) and (**c**) represent, respectively, the transfer characteristics of p-channel and n-channel behavior under light illumination. (**d**) and (**e**) are the respective output properties of the device under light condition. (**f**) Output curves of the phototransistor in the dark condition. Solid and dashed curves were measured during forward and backward sweepings, respectively.

Figure 2b and 2c show the plots of $I_{DS}^{1/2}$ and $I_{DS}$ as functions of $V_{GS}$ for the perovskite phototransistor measured under light illumination. The plots appear linear for a large range of $V_{GS}$ for both p-type and n-type transports. The field-effect mobility ($\mu$) and threshold voltage ($V_{TH}$) can be extracted using the saturated $I_{DS}$ versus $V_{GS}$ relationship[52,53]:

$$I_{DS} = \frac{W}{2L} C_i \mu (V_{GS} - V_{TH})^2 \quad (1)$$

where $W$, $L$, and $C_i$ are the channel width, the channel length, and the gate capacitance per unit area, respectively. Accordingly, the photo-induced hole and electron mobilities are derived as 0.18 $cm^2\ V^{-1}\ s^{-1}$ and 0.17 $cm^2\ V^{-1}\ s^{-1}$, respectively, in the saturation region. Thus, it is clear that under light illumination the perovskite-based phototransistor exhibits ambipolar behavior with balanced hole and electron mobilities.

Under light illumination, the output characteristics (i.e., $I_{DS}$ versus $V_{DS}$ data at different $V_{GS}$) obtained for the device operating in the hole-enhancement and electron-enhancement modes are displayed in Fig. 2d and 2e, respectively. At low $V_{GS}$, the device exhibits ambipolar transport with diode-like current-voltage ($I$-$V$) characteristics; however, at high $V_{GS}$, unipolar transport with standard linear-to-saturation $I$-$V$ transistor characteristics was observed. Figure 2f shows the output characteristics of the phototransistor measured in the dark. It can be clearly seen that there is no obvious field effect, and $I_{DS}$ remains at the level of $10^{-10}$ A, which is consistent with the transfer characteristics.

For measurements performed in the dark, as shown in Fig. 2a, $I_{DS}$ remains less than 0.5 nA when $V_{GS}$ increases from −40 V to +40 V, and the obtained On/Off ratio is less than 10. In contrast, regarding device performance under light illumination, the On current is significantly enhanced to $10^{-4}$ A, while the Off current also increases to $2.0 \times 10^{-9}$ A. Furthermore, the On/Off ratios under illumination are boosted to $3.32 \times 10^4$ for p-type and $8.76 \times 10^3$ for n-type transport. The significant gate tuning effect upon illumination with respect to the dark measurements implies that the photo-excited carriers dominate the channel transport; thus, the device has potential for use in other photoelectronic applications such as photosensors.

Furthermore, it should be noted that both the transfer curves and the output curves exhibit weak but notable hysteresis effects (Fig. 2), which were also reported for solar cells, memristors and light-emitting field-effect transistors based on hybrid perovskites[31,57-59]. One origin of the hysteresis effects was speculated to be the screening effects arisen from the field-induced drift of methylammonium cations[59]. Additionally, charge traps, and surface dipoles at the untreated semiconductor-dielectric interface may also attribute to the hysteresis effect[53]. In Fig. 2, the hysteresis effect in the phototransistor operation is not substantial, which is presumably a result of the high quality of the perovskite films.

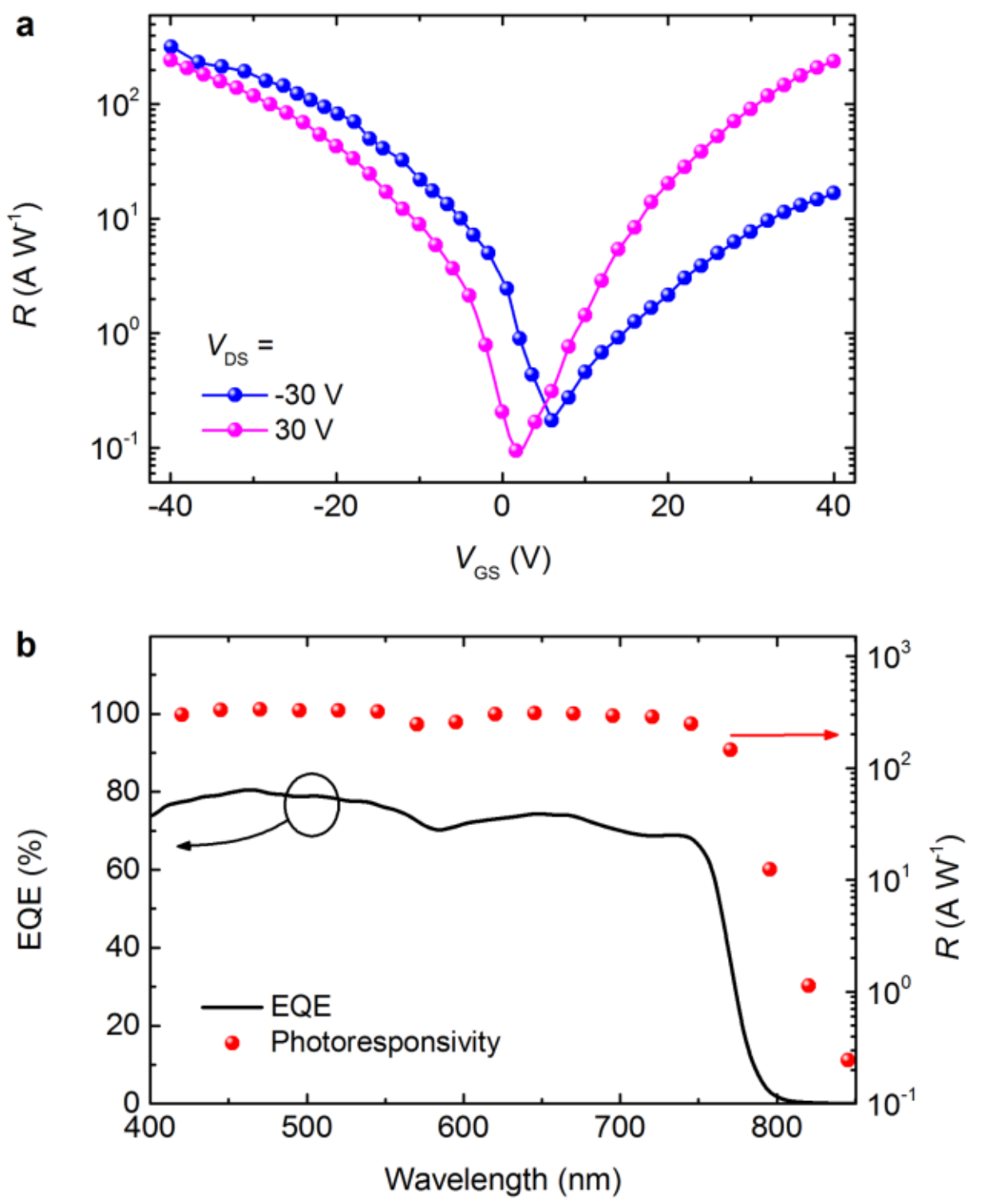

**Figure 3.** (**a**) Photoresponsivity (*R*) characteristics of the $CH_3NH_3PbI_3$ phototransistor under white light illumination (power density: 10 mW cm$^{-2}$) when the drain voltages ($V_{DS}$) are −30 V

and 30 V, respectively. (**b**) External quantum efficiency (EQE) and photoresponsivity (*R*) of the hybrid perovskite photodetector at different wavelength when the drain voltage ($V_{DS}$) is −30 V.

To examine the effect of the device structure, we also fabricated a bottom-gate bottom-contact perovskite phototransistor (schematic shown in Supplementary Figure 2). Supplementary Figure 3 shows the ambipolar transfer characteristics of the devices under light illumination, which is similar to that of the bottom-gate top-contact devices. In this particular device, both the electron and hole mobilities were measured as 0.17 $cm^2 V^{-1} s^{-1}$, suggesting a very balanced ambipolar transport under light illumination. These results reveal that the ambipolar performance of phototransistors is an intrinsic property of $CH_3NH_3PbI_3$ and is independent of the device architecture.

Photoresponsivity (*R*) is the key parameter for evaluating the performance of phototransistors, and its value can be obtained using transfer characteristics. *R* is given by the following equation:

$$R = \frac{I_{light} - I_{dark}}{E_{light}} \quad (2)$$

where $I_{light}$ and $I_{dark}$ stand for the drain currents under light illumination and in the dark, respectively, and $E_{light}$ is the power of incident illumination. *R* as a function of $V_{GS}$ is shown in Fig. 3a. The maximum *R* value of 320 A $W^{-1}$ in the On-state ($V_{GS}$ = −40 V) was obtained at a power density of 10 mW $cm^{-2}$. We also examined the phototransistors with various perovskite film thicknesses (shown in Supplementary Figure 4), and we found that the performance of devices with the active layer thickness of about 100 nm is optimal. Note that this parameter achieved in perovskite phototransistors is substantially higher than those of most other reported functional materials[38,40,42-45,47,48]. For example, phototransistors based on organic semiconductors and hybrid materials were reported with *R* values typically below 1.0 A $W^{-1}$ (reference 45). This superb performance of hybrid perovskites in phototransistors can be attributed to their excellent light absorption properties, extremely low dark current and high photocurrent.

The external quantum efficiency (EQE) and photoresponsivity ($R$) as the function of wavelength in our phototransistor are shown in Fig. 3b. The device exhibits a broad photoresponse range from 400 to 800 nm, and its maximum EQE is approximately 80%. Moreover, the device shows high $R$ values in the wavelength range of 400-750 nm, and hence it is suitable for visible-light broadband photodetection. In addition, we found that the spectral sensitivities of the phototransistor are mainly determined by the absorption spectra of the perovskite film (Fig. 1b).

It was reported that Cl might act as a crystallization retarding and directing agent[60], which benefits the growth of perovskite domains and thus improves the transport properties of the perovskite films[21,22]. Therefore, we fabricated mixed-halide perovskite $CH_3NH_3PbI_{3-x}Cl_x$-based phototransistors. Indeed, we found that the presence of Cl improves the surface smoothness of the perovskite films (Supplementary Figure 5). The transfer characteristics of the $CH_3NH_3PbI_{3-x}Cl_x$-based phototransistor in dark and light conditions are shown in Fig. 4a. Under light illumination (10 mW cm$^{-2}$), the device shows clear ambipolar behavior, similar to that of $CH_3NH_3PbI_3$-based devices. As shown in Fig. 4b and 4c, the slopes of $I_{DS}^{1/2}$ versus $V_{GS}$ are linear for a large range of $V_{GS}$ for both p-type and n-type transport, which demonstrates the high quality of our phototransistors. Based on Equation (1), the obtained mobilities of photo-induced holes and electrons are 1.24 cm$^2$ V$^{-1}$ s$^{-1}$ and 1.01 cm$^2$ V$^{-1}$ s$^{-1}$, respectively. However, unlike the case of $CH_3NH_3PbI_3$ phototransistors, $I_{DS}$ in $CH_3NH_3PbI_{3-x}Cl_x$ devices can reaches 10 nA in the dark condition, and the hole and electron mobilities are $1.62 \times 10^{-4}$ cm$^2$ V$^{-1}$ s$^{-1}$ and $1.17 \times 10^{-4}$ cm$^2$ V$^{-1}$ s$^{-1}$, respectively (Supplementary Figure 6). These results obtained in the dark condition imply that the $CH_3NH_3PbI_{3-x}Cl_x$ channels have substantially higher conductivity and mobility than the $CH_3NH_3PbI_3$ ones. As shown in Supplementary Figure 7, the maximum *R* value of the $CH_3NH_3PbI_{3-x}Cl_x$ phototransistor is approximately 47 A $W^{-1}$ at $V_{GS} = -40$ V. The lower *R* of the $CH_3NH_3PbI_{3-x}Cl_x$ phototransistor compared to that of the $CH_3NH_3PbI_3$ phototransistor is primarily due to its higher dark current. Nevertheless, this *R* value is still comparable with the highest ones reported for other functional materials[38,40,42-45,47,48].

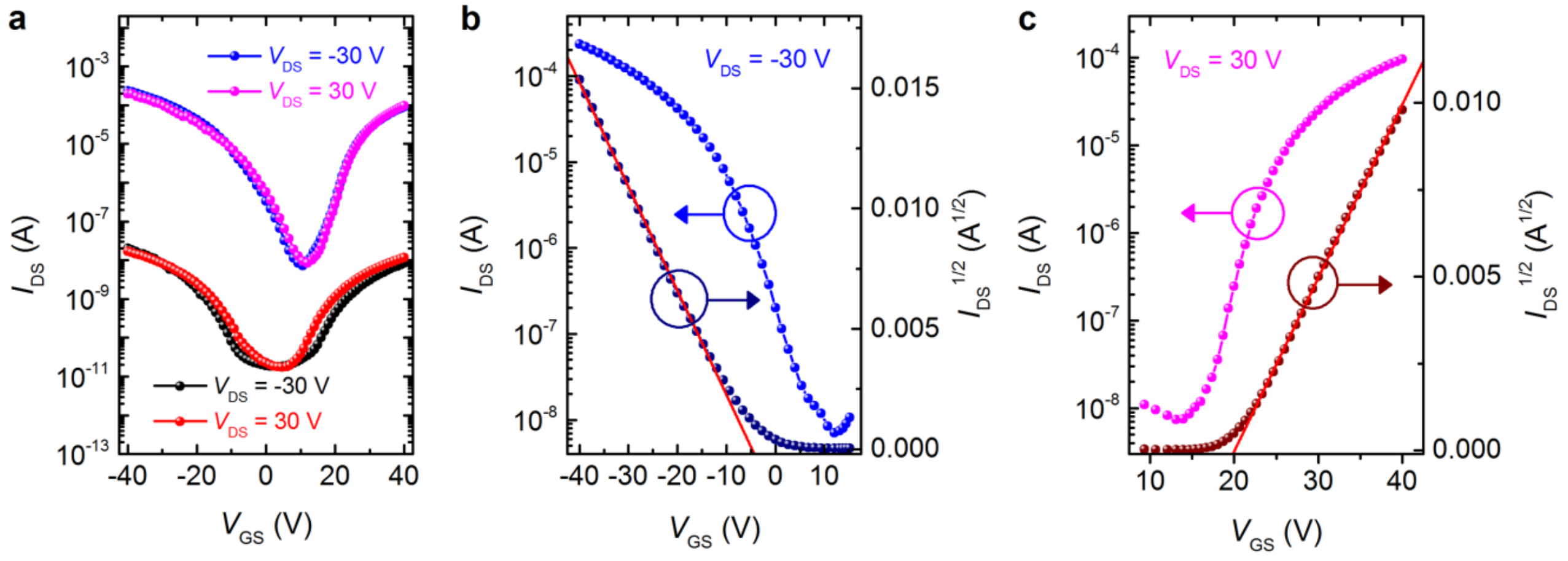

**Figure 4.** (**a**) Transfer curves of the $CH_3NH_3PbI_{3-x}Cl_x$ phototransistor in the dark (black and red dots) and under light illumination (blue and magenta dots). (**b**) and (**c**) represent, respectively, the transfer characteristics of p-type and n-type behaviors under light illumination.

Fast response to optical signals, which is a result of efficient charge transport and collection, is critical for optoelectronic devices. The time-dependent photocurrent was recorded when the white light (10 mW $cm^{-2}$) was turned on and off regularly. Figure 5a shows the response of photocurrent to optical pulses at a time interval of 0.5 s when the device is biased at $V_{DS}$ = -30 V and $V_{GS}$ = -30 V. We found that the dynamic photoresponse of the perovskite phototransistors was stable and reproducible. The photocurrent quickly increases as soon as the light is turned on and then drops to the original value when the light is turned off, indicating that the device functions as a good light-activated switch.

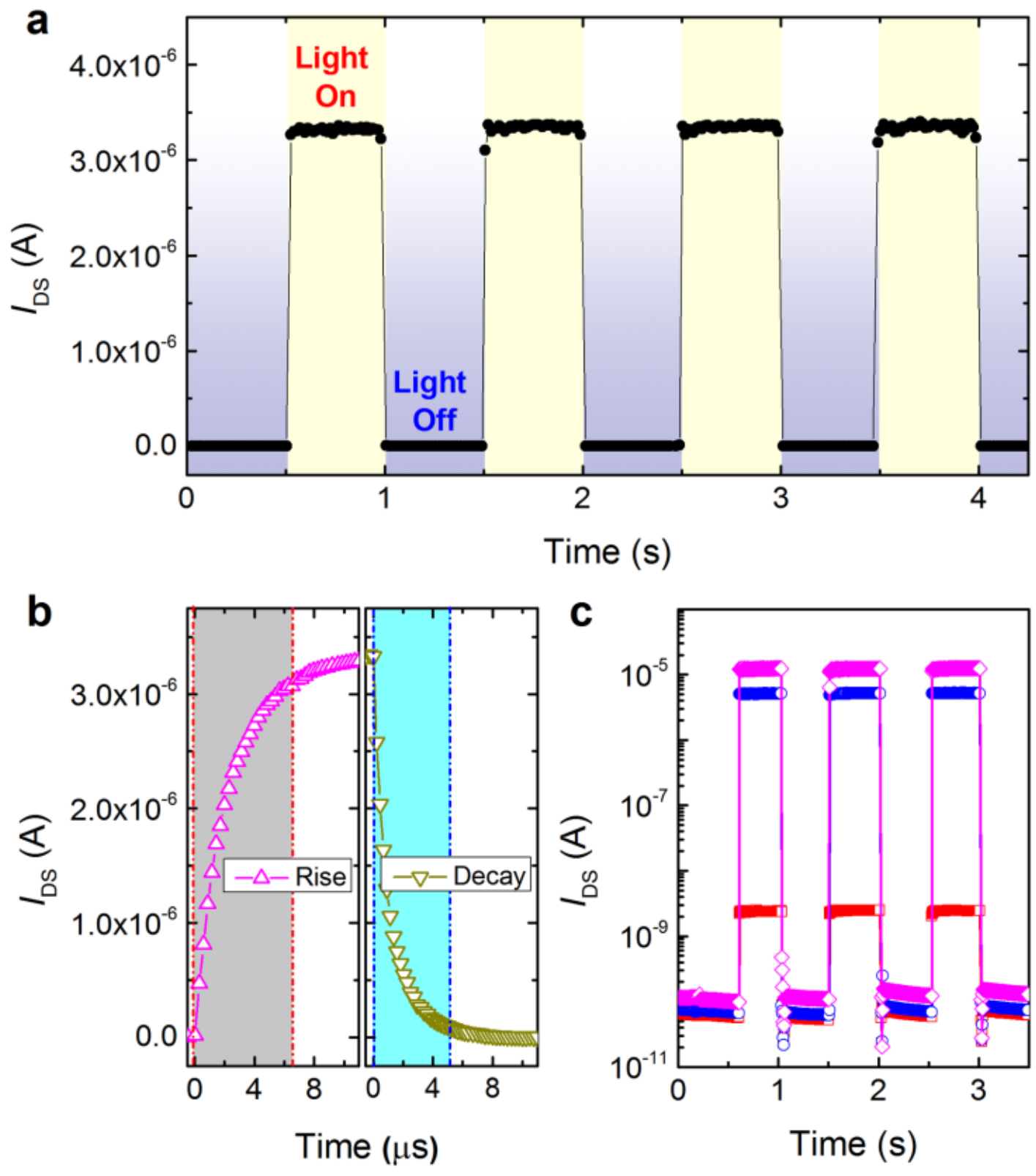

**Figure 5.** (**a**) Photocurrent responses of the phototransistors upon light illumination showing time-dependent photosensitivity with a time interval of 0.5 s at $V_{GS}$ = −30 V and $V_{DS}$ = −30 V. (**b**) Temporal photocurrent responses highlighting a rise time of 6.5 μs and a decay time of 5.0 μs. (**c**) Gate voltage-dependent photoresponses measured at $V_{DS}$ = −30 V. The device was measured at various $V_{GS}$ of 0 V (red square), −30 V (blue circle) and −40 V (magenta diamond), respectively.

Figure 5b shows the temporal photocurrent response of the perovskite phototransistor. The measured switching times for the rise ($I_{DS}$ increasing from 0 to 80% of the peak value) and decay ($I_{DS}$ decreasing from the peak value to 10%) of the photocurrent are approximately 6.5 μs and 5.0 μs, respectively, indicating an ultrafast response speed. The decay time of the photodetector has previously been estimated as the carrier recombination time (lifetime, $\tau_{life}$)[39,41,43]. Recently, using a solar-cell-type photodetectors, Dou *et al.* obtained response times below 200 ns[26]. In fact, photo-induced electron–hole pairs in the pristine perovskite recombine within a few picoseconds (i.e., within the lifetime of the photo-excited electrons)[30]. However, in the presence of positive (negative) gate voltages, the photo-generated electrons (holes) are accumulated in the channel and acquire a higher mobility, while the other type of charges remain trapped in the perovskite channel. Thus, the recombination of the photo-excited electron-hole pairs in the perovskite channel is reduced, resulting in increased carrier lifetime ($\tau_{life}$)[43,61]. Nevertheless, the temporal photocurrent responses of our perovskite phototransistors are still substantially faster than those of most organic, quantum dot and hybrid photodetectors (typically in the order of milliseconds)[36,41-49].

Figure 5c shows the characteristics of photo-switched channel current at $V_{DS}$ = -30 V and various $V_{GS}$. As shown, upon light illumination, a larger $V_{GS}$ can induce a more significant increase of photocurrent. In particular, the photocurrent is approximately $2.76 \times 10^{-9}$ A at $V_{GS}$ = 0 V, whereas it dramatically increases to $1.21 \times 10^{-5}$ A at $V_{GS}$ = -40 V. These results demonstrate that the switching of the photocurrent could be tailored to a large degree using the gate voltages, enabling highly tunable photodetection.

To enhance the stability of perovskite phototransistors, we coated some bottom-gate top-contact devices with poly(methyl methacrylate) (PMMA) protective layers (noted as Device-P). These devices were stored in ambient conditions without isolation from exposures to atmospheric air and moisture. Previous works on perovskite solar cells have demonstrated that coating the PMMA protective layer is an effective strategy to improve the ambient stability of perovskites by forming kinetic barriers against the diffusion of moisture and/or atmospheric oxygen[21]. Under light illumination, the transfer curves of a device immediately following fabrication and after storing in ambient conditions for 60 days are essentially the same except slight decreases of the channel current (shown in Fig. 6a), revealing excellent device stability. In contrast, the unprotected devices without the PMMA layer (noted as Device-U) exhibited sustantially weaker stability. Figure 6b shows the transfer characteristics of a newly fabricated Device-U, which are similar to those of Device-P. The inset of Fig. 6b shows the transfer curve of Device-U that had been stored in ambient conditions for 60 days. Notably, the device showed $I_{DS}$ of only approximately $10^{-8}$ A at $V_{GS}$ = -40 V, which is nearly $10^{4}$-fold lower than the $I_{DS}$ of the newly fabricated Device-U. Figure 6c shows the mobility and On/Off ratio of the Device-P measured at different intervals over a period of 60 days. No apparent statistical variations in the mobilities or On/Off ratio were observed, confirming the excellent stability of the encapsulated phototransistors. For the Device-U (shown in the inset of Fig. 6c), both the mobility and the On/Off ratio declined continuously during 60 days of storage. However, it should be mentioned that the performance of Device-U changed very little over the first 7 days, indicating that unprotected devices must be measured quickly in order to achieve reproducible data.

High reproducibility and stability are critical for photoelectronic devices and their integrations in real-world applications. Figure 6d shows the distributions of hole mobility for both PMMA-protected and -unprotected as-fabricated devices. More than 70% of the Device-P group (20 devices) show hole mobility ranging from 0.13 $cm^2 V^{-1} s^{-1}$ to 0.19 $cm^2 V^{-1} s^{-1}$. However, among the 20 devices in the Device-U group, there are notable fluctuations of mobility values: some are similar to Device-P, while others are even below $9 \times 10^{-2}$ $cm^2 V^{-1} s^{-1}$. The perovskite phototransistors without the PMMA layer often exhibit low reproducibility, primarily due to their high sensitivity toward moisture, room light, and other factors in ambient conditions. Thus, coating PMMA protective layers should be considered as a general protocol for enhancing device stability and reproducibility.

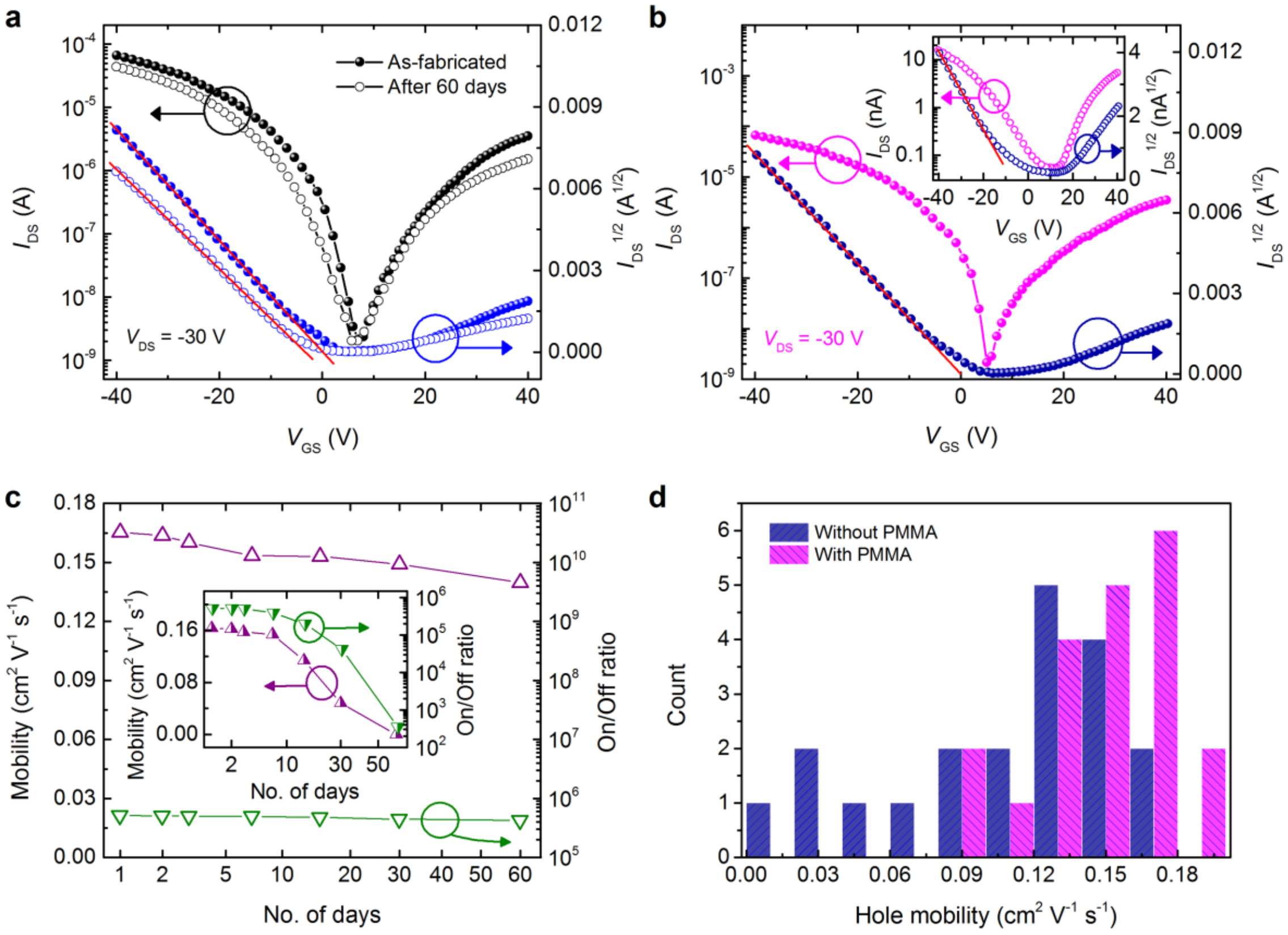

**Figure 6.** (**a**) Transfer curves under light condition of the Device-P (with PMMA protective layer) immediately following fabrication and after storing in ambient conditions for 60 days. (**b**) Transfer characteristics under light condition of the as-fabricated Device-U (with PMMA protective layer) and the data after storing in ambient conditions for 60 days (shown in the inset). (**c**) Mobility and On/Off ratio of Device-P measured during 60 days of storage in ambient conditions. Stability data of Device-U are also shown in the inset for comparison. (**d**) Histogram of the hole mobilities measured on 20 devices with the PMMA protective layer and 20 devices without the protective layer.

In this work, ambipolar phototransistors based on solution-processed hybrid perovskites were fabricated and characterized. Thanks to the superior optical and electronic properties of hybrid perovskites, our solution-processed phototransistors demonstrated high performances and tunability. Under light illumination, these devices exhibit reliable high performance with balanced carrier mobilities and On/Off ratios of approximately $10^4$ for both p- and n-type transports. Remarkably, the phototransistors show excellent figures-of-merit such as excellent photoresponsivity (320 A $W^{-1}$) and ultrafast response speed (less than 10 μs).

In such phototransistors, both light illumination and gate bias can be used to modulate the transport of semiconducting perovskite channels. Through the capacitive coupling, the gate bias is expected to effectively separate the photo-generated holes and electrons, increasing their recombination time, equivalently $\tau_{\mathrm{life}}$ (references 43, 61). The photoconductive gain ($G$), i.e., the number of carriers collected per photo-induced carrier, is given by the equation:

$$G = \frac{\tau_{\mathrm{life}}}{\tau_{\mathrm{tran}}} = \frac{\tau_{\mathrm{life}}}{L^2 / \mu \cdot V_{\mathrm{DS}}} \quad (3)$$

where $\tau_{\mathrm{tran}}$ is the carrier transit time[39,41-43]. For our devices with mobility in the range of 0.2 - 2.0 $cm^2\,V^{-1}\,s^{-1}$, $G$ is estimated to be approximately 10 - $10^2$. Through biasing the gate terminal, our phototransistor architecture is advantageous for enhancing the photoresponse.

Furthermore, according to Equation (3), the device performance should also benefit from high carrier mobility ($\mu$). In our phototransistors, the field-effect hole/electron mobility was measured as 0.18/0.17 $cm^2\,V^{-1}\,s^{-1}$ for $CH_3NH_3PbI_3$ and 1.24/1.01 $cm^2\,V^{-1}\,s^{-1}$ for $CH_3NH_3PbI_{3-x}Cl_x$, this is consistent with the Cl-enhanced electron-hole diffusion lengths observed in previous works on perovskite solar cells[21,22]. However, better phototransistor performance was achieved in the undoped $CH_3NH_3PbI_3$ channel as a result of suppressed dark current. We should note that although the field-effect mobility measured in our perovskite phototransistors is impressive for solution-processes materials, there is still room for improvements. Tailoring the deposition procedures and developing appropriate doping strategies are promising to optimizing the transport properties of perovskite channels[62,63].

Furthermore, according to Equation (3), reducing the channel length ($L$) is a straightforward way to improve the performance of phototransistors. In the present device structure, the channel length is 50 μm, which are much longer than the diffusion lengths of the carriers. Therefore, light-induced carriers are scattered many times by structural defects and grain boundaries before reaching the electrodes[64,65]. Much higher carrier mobility is expected in phototransistors with shorter channel lengths, which could further improve the phototransistor performance. Shrinking the channel dimensions may also help decrease the $V_{DS}$ required for the device operation and reduce the energy consumption. Overall, our findings strongly support the use of hybrid perovskites as active materials in high-performance ambipolar phototransistors, and open new doors for employing such solution-processed perovskites in other photoelectronic applications.

## Methods

**Perovskite preparation and device fabrication.** $CH_3NH_3I$ was synthesized according to a previously reported procedure[2,6,11]. First, 24 mL of methylamine (33 wt% in absolute ethanol, Sigma) and 10 mL of hydroiodic acid (57 wt% in water, Aldrich) were mixed to react in a 250 mL round-bottomed flask at 0 °C for 2 h with stirring. The precipitate was recovered via evaporation at 50 °C for 1 h. The product, methyl ammonium iodide $CH_3NH_3I$, was washed with diethyl ether by stirring the solution for 30 min, which was repeated three times. The product was finally dried at 60 °C in a vacuum oven for 24 h. The perovskite films were prepared according to the vapor-assisted solution process[11]. The heavily n-doped Si wafers with 300-nm $SiO_2$ surface layers (capacitance of 15 nF $cm^{-2}$) were used as the substrates. They were successively cleaned with diluted detergent, rinsed with deionized water, acetone and ethanol, and dried with dry nitrogen. After the oxygen plasma treatment, the solution of 0.3 M $PbI_2$ or $PbCl_2$ (Sigma) in dimethylformamide (DMF) was spin-coated on the cleaned Si substrates at 5000 r.p.m. for 40 s and dried at 110 °C for 15 min. $CH_3NH_3I$ powders were spread around the $PbI_2$ or $PbCl_2$ coated substrates, and a petridish was placed over the samples. The substrates were heated at 150 °C for 8 hours. After cooling, the as-prepared films were washed with isopropanol and dried at 65 °C for 5 min. Finally, Ti/Au (5 nm/80 nm) source (S) and drain (D) electrodes were deposited via thermal evaporation through a shadow mask, defining a channel length (*L*) of 50 μm and a channel width (*W*) of 1000 μm. Furthermore, the fabricated devices were annealed in a glove box at 50-60 °C for 10 min to reduce the charge traps and improve the contact between the active layer and the S/D electrodes[66]. The bottom-gate bottom-contact devices were fabricated following similar procedures except that Au S/D electrodes were deposited on the Si substrates prior to spin-coating the $PbI_2$ or $PbCl_2$ solution. To improve device stability, thin layers of PMMA (average MW approximately 350000 by GPC, Sigma-Aldrich) were spin-coated onto the surfaces of the perovskite films at 8000 rpm for 60 s.

**Measurements and experimental set-up.** Perovskite film thickness was assessed using the scratch method via atomic force microscopy (AFM, Bruker Dimension ICON). X-ray diffraction (XRD) patterns of the perovskite films were recorded using a Bruker D8 ADVANCE diffractometer with Cu $K\alpha$ ($\lambda$ = 1.5406 Å) radiation. A field-emission scanning electron microscope (FESEM, FEI Nova Nano 630) was used to acquire surface SEM images. Additionally, the morphology of the perovskite films was determined via AFM measurements in the tapping mode. The absorption spectra were measured using an Agilent Cary 6000i UV-Visible-NIR spectrometer. The devices current-voltage (*I-V*) measurements were performed using a Keithley 4200 Semiconductor Parametric Analyzer and a Signotone Micromanipulator S-1160 probe station. A light-emitting diode (white light, 10 mW $cm^{-2}$) attached to the microscope of the probe station was used as the light source. During the measurements, the samples were kept at room temperature in the ambient atmosphere. The external quantum efficiency (EQE) of the device was measured using an Oriel IQE-200 measurement system.

## Acknowledgements

This work was supported by the King Abdullah University of Science and Technology (KAUST).

# Supplementary Information

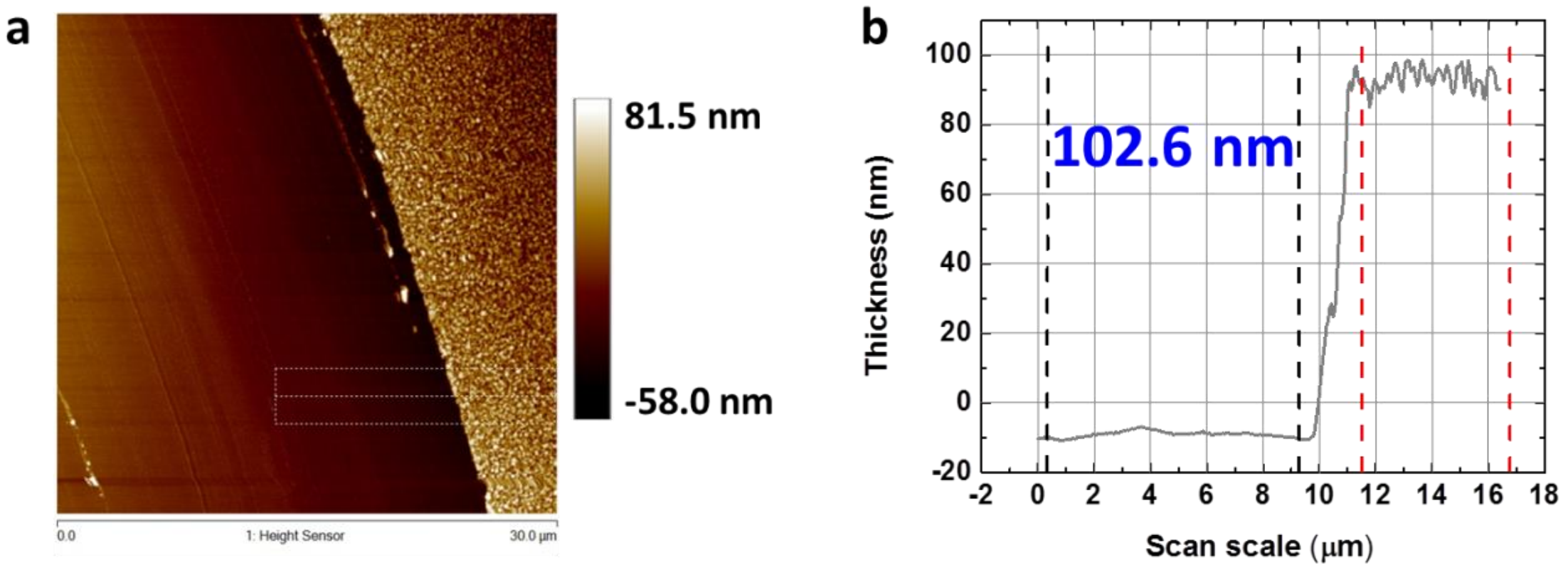

**Supplementary Figure S1.** (**a**) AFM images of the perovskite film with a straight edge which was created via scratching with a tweezer. (**b**) Profile along the line highlighted in (**a**). This profile was used to estimate the thickness of the perovskite film.

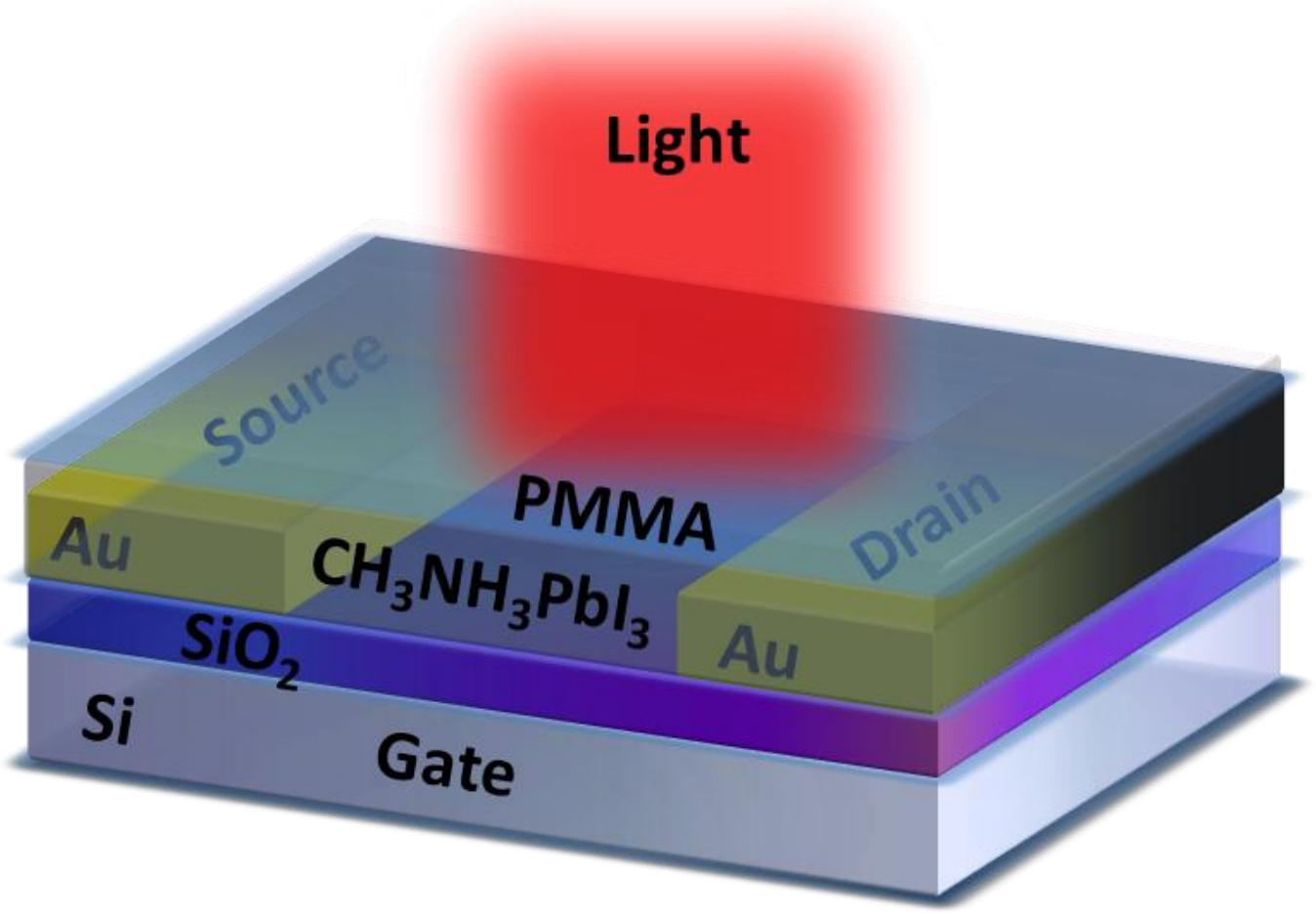

**Supplementary Figure S2.** Schematic diagram for a $CH_3NH_3PbI_3$-based phototransistor with the bottom-gate bottom-contact structure.

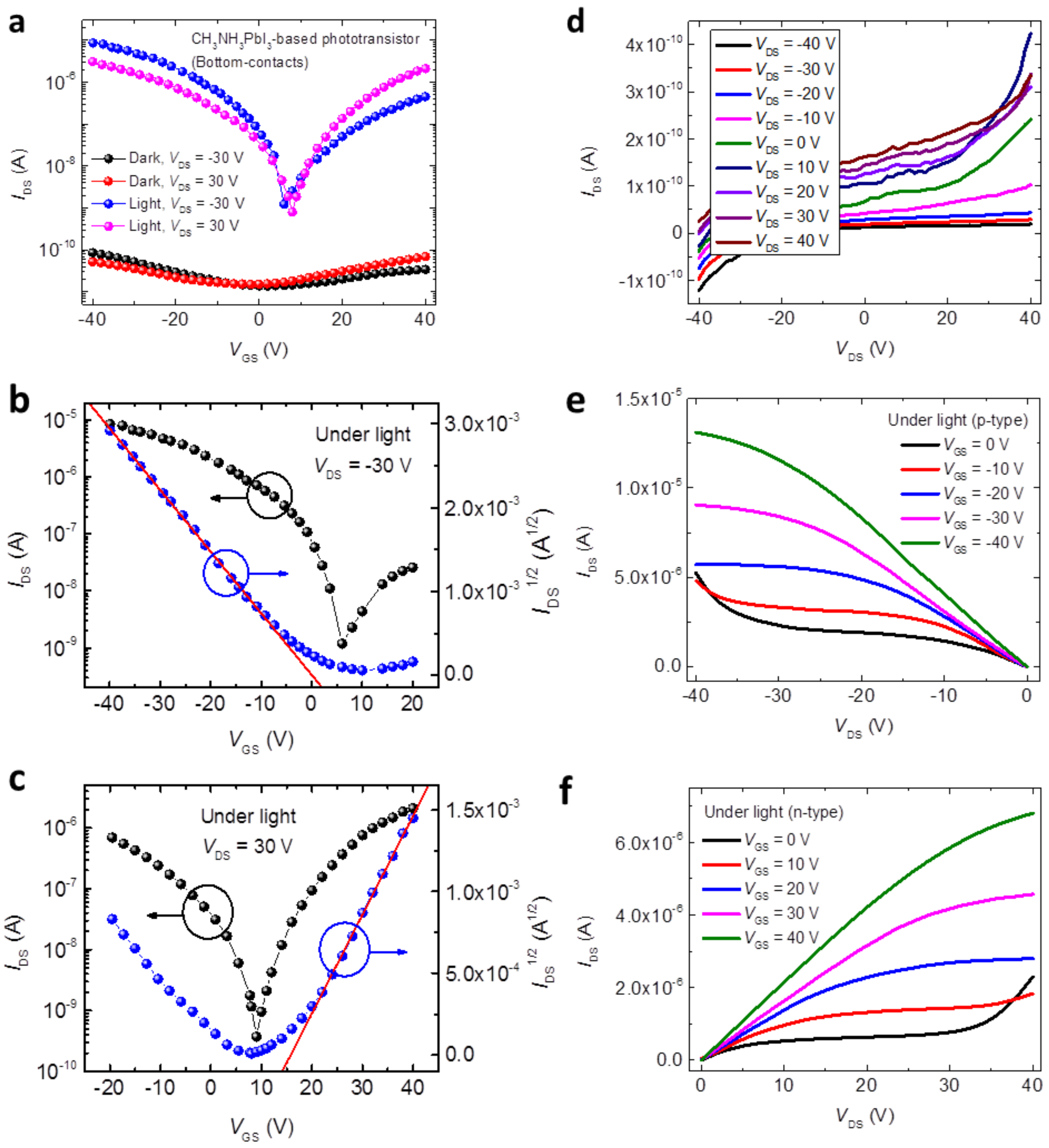

**Supplementary Figure S3.** (**a**) Transfer curves for the bottom-gate, bottom-contact phototransistor device. (**b**) and (**c**) represent, respectively, the transfer characteristics of p-type behavior and n-type behavior. (**d**), (**e**) and (**f**) are output curves of the phototransistor under dark condition and under light illumination.

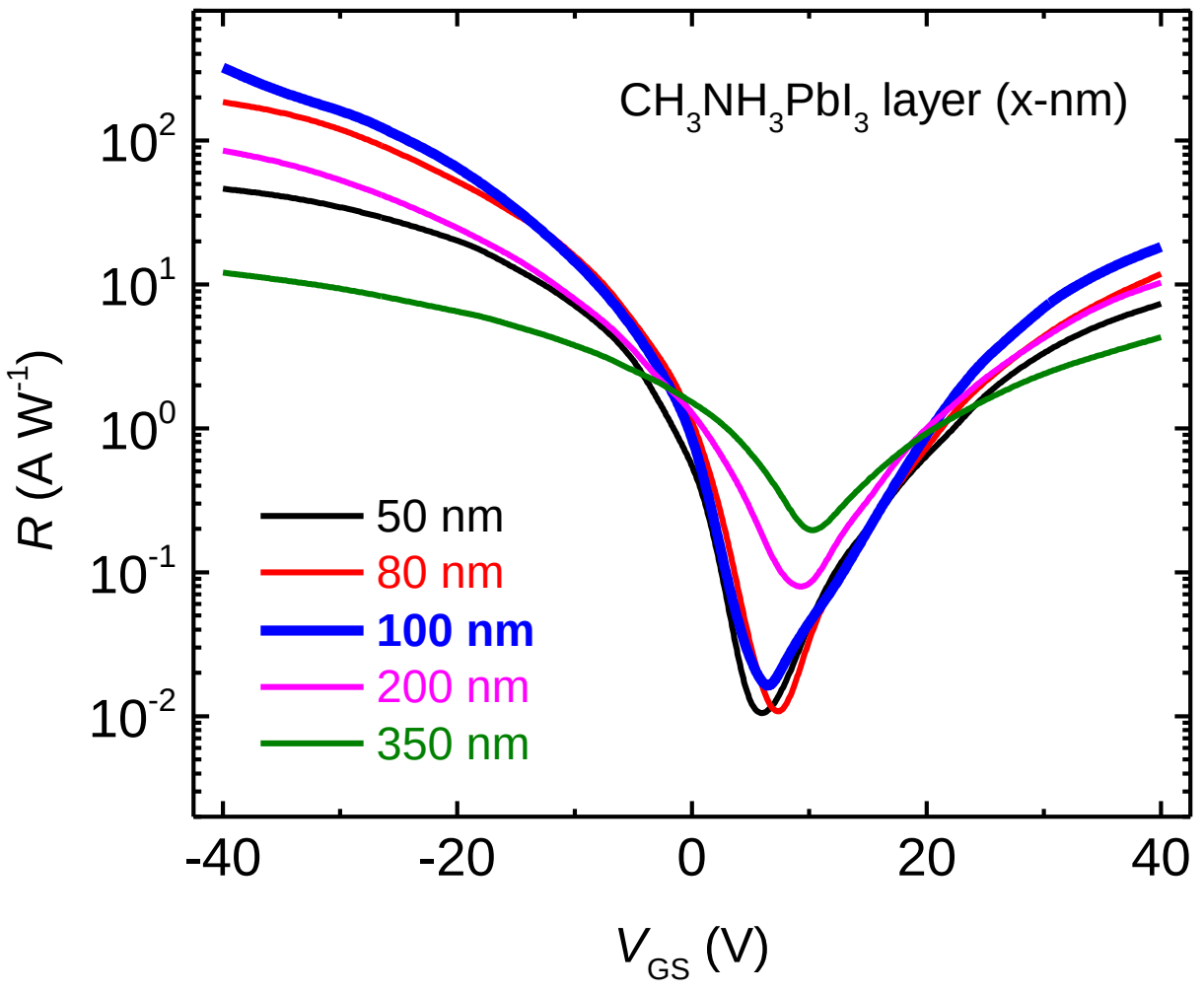

**Supplementary Figure S4.** The device with a thickness of about 100 nm was found to present the optimal performance.

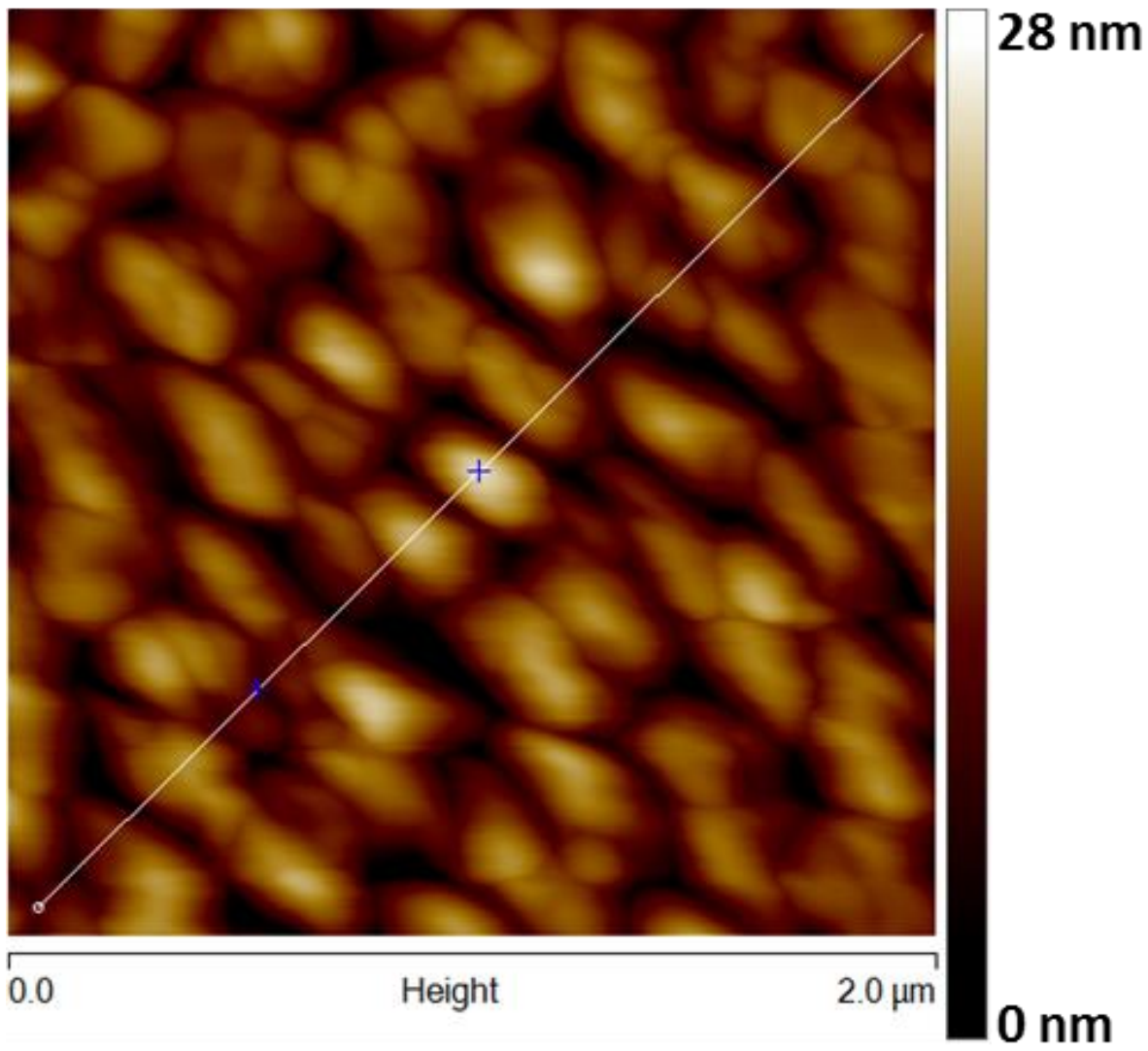

**Supplementary Figure S5.** The root-mean square roughness is approximately 8.95 nm, demonstrating a smoother surface compared to the $CH_3NH_3PbI_3$ film (shown in Fig. 1d in the main text).

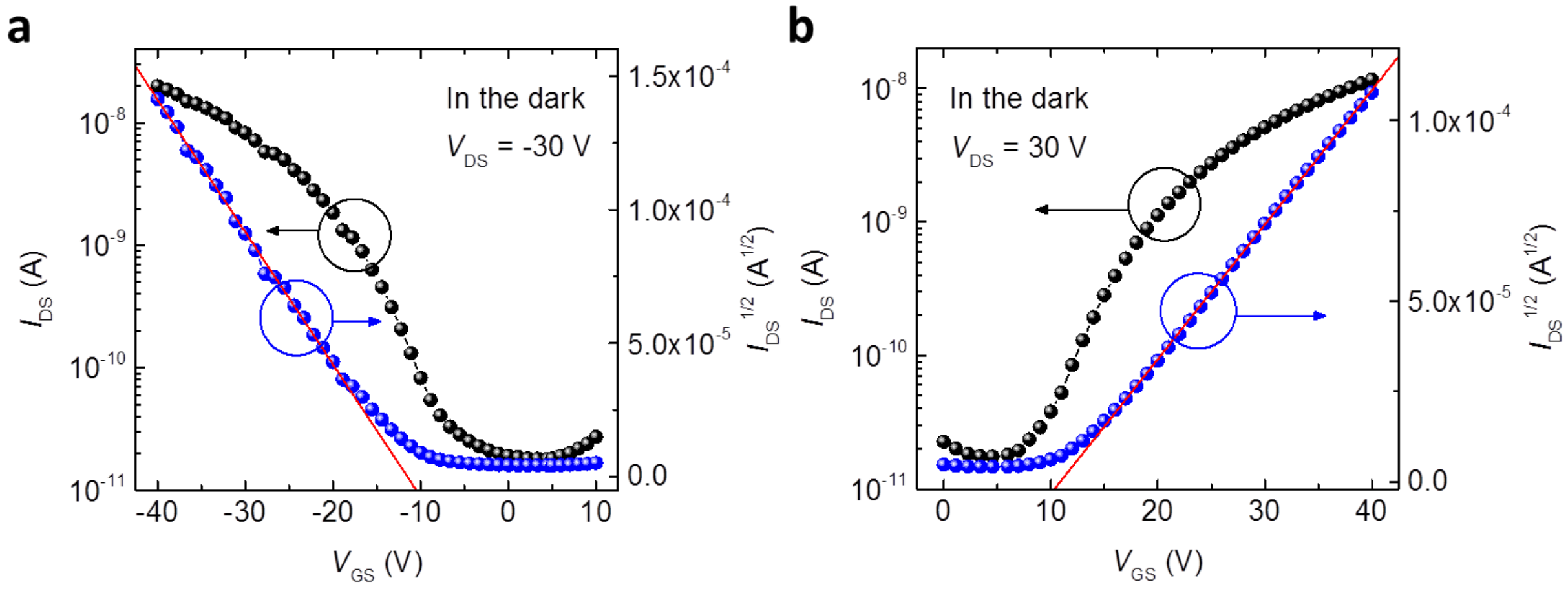

**Supplementary Figure S6.** (**a**) and (**b**) represent, respectively, the transfer characteristics of a hybrid perovskite $CH_3NH_3PbI_{3-x}Cl_x$-based phototransistor in the dark with p-type behavior and n-type behavior.

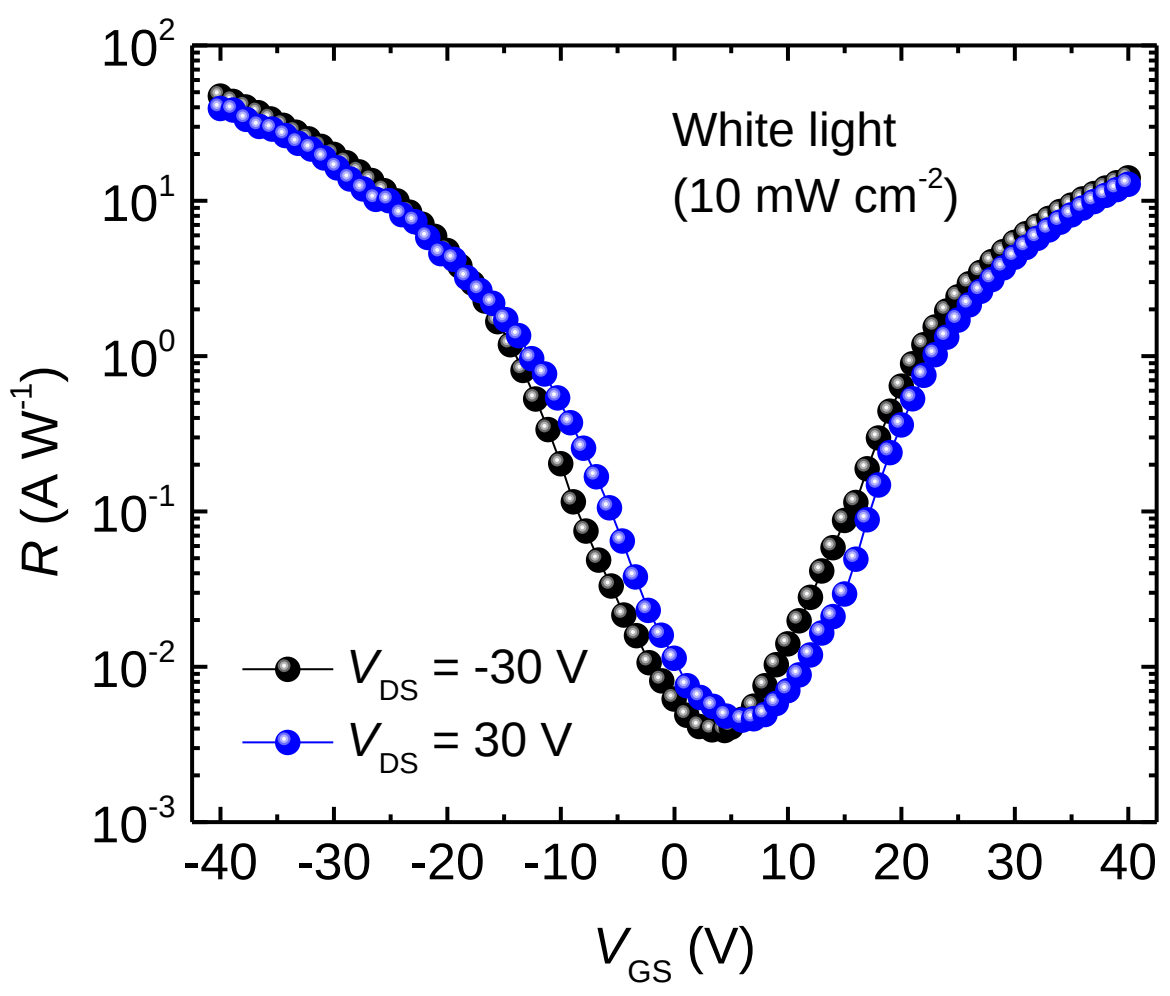

**Supplementary Figure S7.** Photoresponsivity (*R*) values for the perovskite $CH_3NH_3PbI_{3-x}Cl_x$-based phototransistor when the drain voltages are −30 V and 30 V, respectively.

**Supplementary Table S1. Performance of reported photodetectors based on hybrid perovskites.**

| Ref. | Year | Materials | Configuration | Responsivity ($A\ W^{-1}$) | Detectivity (Jones) | Response time |
|---|---|---|---|---|---|---|
| 1 | 2014 | $CH_3NH_3PbI_3$ film | solar cell | Photocurrent amplification > 100 | | |
| 2 | 2014 | $CH_3NH_3PbI_3/TiO_2$ film | photodetector | $0.49 \times 10^{-6}$ | | 0.02 s |
| 3 | 2014 | $CH_3NH_3PbI_3$ film | photodetector | 3.49 | | < 0.2 s |
| 4 | 2014 | $CH_3NH_3PbI_{3-x}CI_x$ film | photodetector | | $\sim 10^{14}$ | 160 ns |
| 5 | 2014 | $CH_3NH_3PbI_3$ nanowires | phototransistor | $5 \times 10^{-3}$ | | < 500 μs |
| 6 | 2015 | $CH_3NH_3PbI_3$ film | photodetector | 14.5 | | 0.2 μs |
| 7 | 2015 | Graphene- $CH_3NH_3PbI_3$ composites | phototransistor | 180 | $\sim 10^{9}$ | 87 ms |
| 8 | 2015 | $CH_3NH_3PbI_3$ film | photodetector | 242 | | 5.7 ± 1.0 μs |
| 9 | 2015 | $CH_3NH_3PbI_3$ film | photodiode | | $3 \times 10^{12}$ | < 5 μs |
| 10 | 2015 | $CH_3NH_3PbI_3$ film | photodetector | | $7.4 \times 10^{12}$ | 120 ns |
| 11 | 2015 | $CH_3NH_3PbI_3$ film | optocoupler | 1.0 | | 20 μs |
| 12 | 2015 | $CH_3NH_3PbI_3$ nanowires | photodetector | 1.3 | $2.5 \times 10^{12}$ | 0.3 ms |

## Supplementary References